\documentclass[12pt]{article}
\usepackage{amsmath,amssymb,graphicx}
\usepackage{epsf}
\usepackage{pstricks}

\newcommand{\beq}{\begin{eqnarray}}
\newcommand{\eeq}{\end{eqnarray}}

\newcommand{\centeron}[2]{{\setbox0=\hbox{#1}\setbox1=\hbox{#2}\ifdim
                            \wd1>\wd0\kern.5\wd1\kern-.5\wd0\fi \copy0
                            \kern-.5\wd0\kern-.5\wd1\copy1\ifdim\wd0>\wd1
                            \kern.5\wd0\kern-.5\wd1\fi}}
\newcommand{\ltap}{\>\centeron{\raise.35ex\hbox{$<$}}
                    {\lower.65ex\hbox{$\sim$}}\>}
\newcommand{\gtap}{\>\centeron{\raise.35ex\hbox{$>$}}
                    {\lower.65ex\hbox{$\sim$}}\>}

\newcommand\ZZ{\hbox{\zfont Z\kern-.4emZ}}
\font\zfont = cmss10 
\newcommand{\sfrac}[2]{{\textstyle\frac{#1}{#2}}}

\def\tv#1{\vrule height #1pt depth 5pt width 0pt}

\begin{document}
\begin{titlepage}
\begin{flushright}
{\tt hep-ph/0308038} \\
Saclay t03/112\\
\end{flushright}

\vskip.5cm

\begin{center}
{\huge \bf Towards a Realistic Model of}
\vskip.1cm
{\huge \bf  Higgsless Electroweak Symmetry Breaking}
\vskip.1cm
\end{center}
\vskip0.2cm

\begin{center}
{\bf
{Csaba Cs\'aki}$^{a}$,
{Christophe Grojean}$^{b}$,
\\
{Luigi Pilo}$^{b}$,
{\rm and}
John Terning$^{c}$}
\end{center}
\vskip 8pt

\begin{center}
$^{a}$ {\it Newman Laboratory of Elementary Particle Physics\\
Cornell University, Ithaca, NY 14853, USA } \\
\vspace*{0.1cm}
$^{b}$ {\it Service de Physique Th\'eorique,
CEA Saclay, F91191 Gif--sur--Yvette, France} \\
\vspace*{0.1cm}
$^{c}$ {\it Theory Division T-8, Los Alamos National Laboratory, Los
Alamos,
NM 87545, USA} \\
\vspace*{0.3cm}
{\tt  csaki@mail.lns.cornell.edu,  grojean@spht.saclay.cea.fr,
pilo@spht.saclay.cea.fr, terning@lanl.gov}
\end{center}

\vglue 0.3truecm

\begin{abstract}
\vskip 3pt
\noindent We present a 5D gauge theory in warped space based on a bulk
$SU(2)_L\times SU(2)_R\times U(1)_{B-L}$ gauge group where the
gauge symmetry is broken by boundary conditions. The
symmetry breaking pattern and  the mass
spectrum resembles that in the standard model (SM).
To leading order in the warp factor the $\rho$ parameter and the
coupling of the $Z$ (or equivalently the $S$-parameter) are as in the
SM, while corrections are expected at the level of a percent. 
From the AdS/CFT point of view the model presented here can be viewed
as the AdS dual of a (walking) technicolor-like theory, in the sense that it 
is the presence
of the IR brane itself that breaks electroweak symmetry, and not a
localized
Higgs on the IR brane (which should be interpreted as a composite
Higgs model).  This model predicts the lightest $W$,  $Z$ and $\gamma$ resonances  to be at around 1.2 TeV, and no fundamental (or composite) Higgs particles.

\end{abstract}

\end{titlepage}

\newpage

\setcounter{equation}{0}
\setcounter{footnote}{0}

The last unresolved mystery of the standard model (SM) of
particle physics is the mechanism for electroweak symmetry breaking
(EWSB).
Within the SM it is assumed that a fundamental Higgs scalar is
responsible for
EWSB. This particle has not been observed yet, and its presence
raises other fundamental issues like the hierarchy problem
(that is how to avoid large quantum corrections to the mass of a light
scalar). Nevertheless, the presence of such a Higgs scalar seems to be
necessary, otherwise the scattering amplitudes of the longitudinal
components of the massive $W$ and $Z$ bosons would blow up at scales of
order 1~TeV, indicating new strongly interacting physics.

Recently, in collaboration with H.~Murayama, we have re-examined~\cite{CGMPT}
   the issue of longitudinal gauge boson scattering
and found that there might be an
alternative way to unitarize the gauge boson scattering amplitudes
without a
Higgs, if there is a tower of massive Kaluza-Klein (KK) gauge bosons
present
in these theories (see also~\cite{otherunitarity,SonStephanov} and, for similar considerations in gravitational theories, see~\cite{Schwartz}).
In~\cite{CGMPT} we have presented a toy model
implementing this idea based on an $SU(2)_L\times SU(2)_R\times U(1)_{B-L}$
gauge symmetry in an extra dimension where the gauge symmetry is
broken by boundary conditions (BC's). There we found that the gauge
boson
spectrum somewhat resembles that in the SM, however the $\rho$ parameter
deviated from unity by as much as ten percent, and the lowest
KK excitations of the $W$ and $Z$ were too light for the model to be
considered realistic.

In this paper we consider a similar model in a warped Randall-Sundrum 
(RS)~\cite{RS} extra dimensional
background. The motivation for considering this modification
comes from the AdS/CFT correspondence~\cite{holography}. The main
problem with the flat space model was the massive violation of
custodial $SU(2)$ symmetry which is manifested in the large deviation
of $\rho$ from one, therefore one would like to ensure that custodial
$SU(2)$ be maintained to leading order. A possible solution to this 
problem in the context of anti-de Sitter (AdS) space has been recently
pointed out by Agashe et al.~\cite{KaustubhRaman}.
If one
considers an AdS background, then one has a dual interpretation of
the theory in terms of a spontaneously broken conformal field theory (CFT):
the breaking of the conformal invariance is manifested by the presence
of
an infrared (TeV) brane, and the fields localized on the TeV brane
are interpreted as bound states of the CFT. Gauge fields in the bulk
correspond to global symmetries (that are weakly gauged) on the CFT
side.
This means that the $SU(2)_L\times SU(2)_R$ gauge symmetry in the bulk
will ensure the presence of custodial $SU(2)$ on the CFT 
side \cite{KaustubhRaman}. The symmetry breaking pattern
on the
TeV brane is $SU(2)_L\times SU(2)_R\to SU(2)_D$, which is exactly as in
the
SM, and preserves custodial isospin. The main
difference
between this model and other RS models with gauge fields in the bulk
(such as~\cite{KaustubhRaman,ADS,CET}, see also~\cite{RSbulk}) is
that here electroweak symmetry is broken by the presence of the TeV
brane
itself, rather than by a scalar Higgs localized on the TeV brane.
The models with a TeV brane localized Higgs should be interpreted as the
duals of composite Higgs models, where there is a scalar bound state of
the
strongly interacting CFT that is responsible for electroweak symmetry
breaking.
On the other hand, the model under consideration here, where electroweak
symmetry breaking is due to the BC's on the TeV brane, should be
interpreted as the dual of a (walking) technicolor-like theory \cite{walking}, since it is
the
strong dynamics itself (the appearance of the TeV brane) that breaks the
electroweak symmetry. Note, that in the AdS picture one can interpolate between the technicolor and composite Higgs models by dialing the expectation value of a brane localized Higgs field. For very large VEV's the Higgs expels the wave functions and becomes a theory with 
BC breaking of electroweak symmetries corresponding to technicolor, while for small VEV's
one gets the usual RS picture corresponding to a composite Higgs model. 

The $SU(2)_R\times U(1)_{B-L}$ symmetry has to be broken in the UV
to ensure that one has the right electroweak group at low energies. This
can again be achieved by a BC breaking on the Planck brane, but will
have the
effect of giving corrections to electroweak observables. In the limit
when the
warp factor becomes infinitely large (the Planck brane is moved to the
boundary of AdS) these corrections will vanish, but for a finite warp
factor they will be suppressed by the  ${\rm log}$  of the
warp factor. These are relatively
small compared to the flat-space model considered in~\cite{CGMPT}, but
still about the order of the experimental precision of the electroweak
observables. Therefore these corrections may still turn out to
be too large, but this requires a detailed calculation of the
electroweak precision observables including loop corrections
from the relatively light KK excitations (and excluding the SM Higgs
loops)
to decide whether this particular model can be completely realistic or
not. Either way, we consider the fact that the lowest order predictions
reproduce the SM results without a Higgs to be a confirmation that the
ideas presented in~\cite{CGMPT} could perhaps be implemented in a
realistic way.

We want to study the possibility of breaking the
electroweak symmetry $SU(2)_L\times U(1)_Y$
down to  $U(1)_Q$  by BC's without
relying on a Higgs mechanism in the bulk.\footnote{Other
interesting possibilities for EWSB using extra dimensions is to have 
the Higgs be the extra dimensional component of a gauge field, see 
for example~\cite{A5Higgs}, or to have a warped compactification where the 
would-be zero mode for the gauge field is not normalizable~\cite{ST}.} The BC breaking is
equivalent to a Higgs mechanism on the brane in the limit of very large VEV's
for the brane Higgs fields. These large VEV's will repel the wave
functions of massive modes
and ensure that the Higgs decouples from the gauge and matter
fields~\cite{CGMPT}.
As in~\cite{CGMPT} we will
consider a bulk $SO(4)\times U(1)_{B-L}\sim SU(2)_L\times SU(2)_R
\times U(1)_{B-L}$ gauge group (where
$U(1)_{B-L}$ corresponds to gauging  baryon minus lepton number),
except here we will consider the theory
compactified in a warped RS background~\cite{RS}.
We will use the conformally flat metric
\beq
ds^2=  \left( \frac {R}{z} \right)^2   \Big( \eta_{\mu \nu} dx^\mu dx^\nu - dz^2 \Big)
\eeq
where $z$ is on the
    interval $[R,R^\prime]$.  In RS-type models $R$ is typically
$\sim 1/M_{Pl}$ and
    $R^\prime \sim  {\rm TeV}^{-1}$.  On the TeV brane at $z=R^\prime$
we break $SO(4)$ down to $SU(2)_D$  by Neumann and Dirichlet BC's.
As discussed above, this corresponds, from the AdS/CFT
point of
view, to breaking the $SU(2)_L\times SU(2)_R$ global symmetry of
the CFT to the diagonal subgroup, just as would happen in technicolor
models. On the Planck brane, $z=R$,
we break $SU(2)_R\times U(1)_{B-L}$ down to the usual  $U(1)_Y$
hypercharge again by Neumann and Dirichlet BC's, to ensure that the
low-energy
gauge group without electroweak symmetry breaking is $SU(2)_L\times
U(1)_Y$.
    Thus in the end only $U(1)_Q$, corresponding to electromagnetism,
remains unbroken.
We denote by
$A^{R\,a}_{M}$, $A^{L\,a}_{M}$ and $B_M$ the gauge bosons of
$SU(2)_R$, $SU(2)_L$ and $U(1)_{B-L}$ respectively; $g_5$ is the gauge
coupling
of the two $SU(2)$'s and $\tilde g_5$, the gauge coupling of
$U(1)_{B-L}$.
We impose the following BC's:
\begin{eqnarray}
&
{\rm at }\  z=R^\prime:
&
\left\{
\begin{array}{l}
\partial_z (A^{L\,a}_\mu +A^{R\,a}_\mu)= 0, \
A^{L\,a}_\mu  - A^{R\,a}_\mu =0, \
\partial_z B_\mu = 0,
\\
\tv{15}
(A^{L\,a}_5 +A^{R\,a}_5) = 0, \
\partial_z (A^{L\,a}_5 -A^{R\,a}_5)  = 0, \
B_5 = 0.
\end{array}
\right.
\label{bc1}\\
&
{\rm at }\ z=R:
&
\left\{
\begin{array}{l}
\partial_5 A^{L\,a}_\mu=0, \
    A^{R\,1,2}_\mu=0,
\\
\tv{15}
\partial_z (g_5  B_\mu + \tilde g_5  A^{R\,3}_\mu )= 0, \
\tilde g_5 B_\mu - g_5 A^{R\,3}_\mu=0,
\\
\tv{15}
A^{L\,a}_5=0, \ A^{R\,a}_5=0, \ B_5 = 0.
\end{array}
\right.
\label{bc2}
\end{eqnarray}

These BC's can be thought of as arising from Higgses
on each brane in the limit
of large VEVs which decouples the Higgs from gauge boson scattering
\cite{CGMPT}.
The Higgs on the TeV brane is a bi-fundamental under the two $SU(2)$'s,
while the Higgs on the Planck brane is a fundamental under $SU(2)_R$
and has charge $1/2$ under $U(1)_{B-L}$ so that
a VEV in the lower component preserves $Y=T_3+B-L$. If one insists on
the
breaking by BC picture, and does not want to think of it as Higgs
fields on
the branes with large VEV's one could ask why the gauge couplings
would
ever appear in the BC's. However, that is clearly dependent on what
normalization is chosen for the gauge fields. Above we have chosen
5D canonical normalization. If we went to the ``more natural''
normalization where the gauge couplings appear only in front of the
gauge kinetic term in the denominator, then the BC's on the Planck brane
$z=R$ would simply be $\partial_z (B_\mu+A_\mu^{R \, 3})=0$, and
$B_\mu=A_\mu^{R\, 3}$. Thus it does not seem unnatural for the gauge
couplings to appear in the expressions for the wave functions once we go
back to a canonically normalized basis.

We will work in unitary gauge (the $\xi\to \infty$ limit of the R$_\xi$
gauge discussed in detail in~\cite{CGMPT}). Since the 5th components
of the gauge fields do not have zero modes, they will all decouple
in the unitary gauge. The Euclidean bulk equation of motion satisfied
by spin-1 fields in AdS space is
\begin{equation}
(\partial_z^2 - \frac{1}{z} \partial_z  +q^2) \psi(z)=0,
\end{equation}
where the solutions in the bulk are assumed to be of the form
$A_\mu (q) e^{-iqx} \psi(z)$. The the KK mode expansion is given by
the solutions to this equation which are
of the form
\beq
	\label{eq:Bwv}
\psi^{(A)}_k(z)=z\left(a^{(A)}_k J_1(q_k z)+b^{(A)}_k Y_1(q_k z)\right)~,
\eeq
where $A$ labels the corresponding gauge boson.
Due to the mixing of the various gauge groups, the KK decomposition is
slightly complicated but it is obtained
by simply enforcing the BC's:
\begin{eqnarray}
           \label{eq:KKB}
B_\mu (x,z) & = & g_5\,  a_0 \gamma_\mu (x)
+   \sum_{k=1}^{\infty} \psi^{(B)}_k (z)  \, Z^{(k)}_\mu (x)\, ,
\\
           \label{eq:KKAL3}
A^{L\, 3}_\mu (x,z) & = &
{\tilde g_5} \, a_0 \gamma_\mu (x)
+   \sum_{k=1}^{\infty} \psi^{(L3)}_k (z) \, Z^{(k)}_\mu (x) \, ,
\\
           \label{eq:KKAR3}
A^{R\, 3}_\mu (x,z) & = &
{\tilde g_5}  \, a_0 \gamma_\mu (x)
+    \sum_{k=1}^{\infty}  \psi^{(R3)}_k (z) \, Z^{(k)}_\mu (x) \, ,
\\
           \label{eq:KKALpm}
A^{L\, \pm}_\mu (x,z) & = &
  \sum_{k=1}^{\infty}  \psi^{(L\pm)}_k (z) \, W^{(k)\, \pm}_\mu (x) \, ,
\\
           \label{eq:KKARpm}
A^{R\, \pm}_\mu (x,z) & = &
  \sum_{k=1}^{\infty} \psi^{(R\pm)}_k  (z) \, W^{(k)\, \pm}_\mu (x)  \, .
\end{eqnarray}
Here $\gamma(x)$ is the 4D photon, which has a flat wavefunction due to
the
unbroken $U(1)_Q$  symmetry, and $W^{(k)\, \pm}_\mu (x)$ and
$Z^{(k)}_\mu (x)$ are the KK towers of the massive $W$ and $Z$
gauge bosons, the lowest of which are supposed to correspond to the
observed $W$ and $Z$.

The equation determining the tower of $W$ masses can be read of
by substituting (\ref{eq:KKALpm}-\ref{eq:KKARpm})
into the BC's, and is given by
\begin{equation}
({R}_0-\tilde{R}_0)(R_1-\tilde{R}_1)+
(\tilde{R}_1-{R}_0)(\tilde{R}_0-{R}_1)
=0
\end{equation}
where the ratios $R_{0,1}$ and $\tilde{R}_{0,1}$ are given by
\begin{equation}
R_i \equiv \frac{Y_i(MR)}{J_i(MR)}, \
\tilde{R}_i \equiv \frac{Y_i(MR')}{J_i(MR')}.
	\label{Wtower}
\end{equation}

To leading order in $1/R$ and for
$\log \left(R^\prime/R \right) \gg1$, the lightest solution for this
equation for the mass of the $W^\pm$'s is
\begin{equation}
M_W^2 = \frac{1}{R^{\prime 2} \log \left(\frac{R^\prime}{R}\right)} \, .
\end{equation}
Note, that this result  does not depend
on the 5D gauge coupling, but only on the scales $R,R'$. Taking $R= 10^{-19}$ GeV$^{-1}$ will fix $R^\prime= 2 \cdot 10^{-3}$ GeV$^{-1}$.

The equation determining the masses of the KK tower for the $Z$ (the states that are mostly 
$A^{L 3}$ or $A^{R 3}$) is given by
\begin{equation}
2\tilde{g}_5^2 ({R}_0-\tilde{R}_1)(\tilde{R}_0-R_1)
=g_5^2\left[({R}_0-\tilde{R}_0)(R_1-\tilde{R}_1)+
(\tilde{R}_1-{R}_0)(\tilde{R}_0-{R}_1)\right]
\label{Ztower}
\end{equation}
The lowest mass of the $Z$ tower is approximately given by
\begin{equation}
M_Z^2  = \frac{g_5^2+2 \tilde g_5^{2}}{g_5^2+ \tilde g_5^{2}}
\frac{1}{R^{\prime 2} \log \left(\frac{R^\prime}{R}\right)}  \, .
\end{equation}
Finally, there is a third tower of states, corresponding to the excited modes of the photon
(the particles that are mostly $B$-type), whose masses are given by
\begin{equation}
R_0=\tilde{R}_0.
\end{equation}
This does not have a light mode (the zero mode corresponding to the massless photon has been
separated out explicitly in (\ref{eq:KKB}-\ref{eq:KKAR3})).

In order to check whether these predictions agree with those of the
SM we need to relate the bulk couplings $g_5,\tilde{g}_5$ to the
effective SM couplings $g,g'$. This has to be done by
introducing matter fields. Locally at the
Planck brane ($z=R$), a
$SU(2)_L\times U(1)_{Y}$ subgroup remains unbroken. We can introduce
matter fields localized on this boundary.
For simplicity consider first a scalar $SU(2)_L$ doublet with a
$U(1)_{B-L}$ charge $q$.
Its interactions with the gauge boson KK modes  are generated through
the localized covariant derivative
\begin{equation}
D_\mu \Phi = \partial_\mu \Phi - \frac{i}{2}
\left(
\begin{array}{cc}
2  {\tilde g_5} q B_{\mu}+ g_5 A^{L\, 3}_\mu & 
g_5(A^{L\, 1}_\mu - iA^{L\,2}_\mu) \\
\tv{15}
g_5 (A^{L\, 1}_\mu + i A^{L\, 2}_\mu) &   
2 {\tilde g_5} q B_{\mu} - g_5 A^{L\, 3}_\mu
 \end{array}
\right)_{|z=R} \Phi  .
\end{equation}
Using the KK decomposition (\ref{eq:KKB})-(\ref{eq:KKARpm}), we evaluate
the gauge fields at the boundary and the scalar covariant
derivative becomes
\begin{eqnarray}
\hspace*{-0.1cm}D_\mu \Phi =
\partial_\mu \Phi
&-& 
ig_5 {\tilde g_5} a_0
\left(
\begin{array}{cc}
q+\sfrac{1}{2} & 0\\
0 & q-\sfrac{1}{2}
\end{array}
\right)
\gamma_\mu \Phi
\nonumber \\
&-&
i \sum_{k=1}^{\infty}
\left(
\begin{array}{cc}
{\tilde g_5} q \psi^{(B)}_k + \sfrac{1}{2}g_5 \psi^{(L3)}_k  & 0\\
0 &  {\tilde g_5} q \psi^{(B)}_k - \sfrac{1}{2}g_5 \psi^{(L3)}_k
\end{array}
\right)_{|z=R}
Z^{(k)}_\mu  \Phi
\nonumber \\
&-& 
\sum_{k=1}^{\infty}
i g_5 \sfrac{1}{\sqrt{2}} \, \psi^{(L\pm)}_{k\, |z=R}
\left(
\begin{array}{cc}
0  & W^{(k)\, +}_\mu \\
W^{(k)\, -}_\mu &  0
\end{array}
\right)
\Phi.
\end{eqnarray}

This needs to be matched to the SM expression of the coupling of 
an $SU(2)_L$ doublet with hypercharge $q$ which is given by
\begin{eqnarray}
\hspace*{-0.1cm}
D_\mu \Phi =
\partial_\mu \Phi -i
\left(
\begin{array}{cc}
\frac{g^2-2qg'^2}{2\sqrt{g^2+g'^2}} Z_\mu+e(q+\frac{1}{2}) \gamma_\mu& 
\frac{g}{\sqrt{2}} \, W^+_\mu \\
\frac{g}{\sqrt{2}} \, W^-_\mu & 
\frac{-g^2-2qg'^2}{2\sqrt{g^2+g'^2}} Z_\mu+e(q-\frac{1}{2}) \gamma_\mu
\end{array}
\right)\Phi .
\end{eqnarray}

To be able to identify the first massive KK gauge bosons $Z^{(1)}$ and
$W^{(1)}$ with the SM $Z$ and $W$ we then need to
determine the gauge boson wavefunctions on the Planck brane and the
integral of the square of the wavefunction
in order to determine the normalization.  To leading order (for $R \gg
R^\prime$) the integrals
are dominated by the region near the Planck brane ($z \sim R$), so in
fact the wavefunctions on the
Planck brane are all that is needed.  More specifically, from the expansion
for small arguments of the Bessel functions appearing in~(\ref{eq:Bwv}), the
wavefunction of a mode with mass
$M  \ll 1/R^\prime$ can be written as~\cite{CET}:
\beq
\psi(z) \approx  
c_0 
+ M^2 \left( c_1 z^2 - \frac{c_0}{2} z^2 \log(z/R)\right) 
+ {\cal O} (M^4 z^4),
\eeq
with $c_0$ at most of order one, and 
$c_1$ at most of order  ${\cal O} ( \log(R^\prime/R))$. Thus
\beq
\int_R^{R^\prime} dz \, \left( \frac{R}{z}\right) \psi(z)^2 =
R\left[c_0^2 \log \left( \frac{R^\prime}{R}\right)+ M^2 c_0 c_1
R^{\prime 2}
-\frac{1}{4} M^2 c_0^2  R^{\prime 2}
 (2 \log \frac{R^\prime}{R}-1) +...\right].
\eeq
As we have seen above, $M^2 = {\cal O} (1/\log(R^\prime/R))$, so in the
leading-log approximation:
\beq
\int_R^{R^\prime} dz \, \left( \frac{R}{z}\right) \psi(z)^2 \approx
R\, c_0^2 \log \left( \frac{R^\prime}{R}\right)~.
\eeq
The boundary conditions on the bulk gauge fields give the following
results for the leading term in the wavefunction for the lightest charged
gauge bosons
\beq
c_0^{(L\pm)} = c_\pm~, \ \ 
c_0^{(R\pm)} \approx  0~,
\eeq
while for the neutral gauge bosons we find in the same approximation
\beq
c_0^{(L3)} \approx - \, c~, \ \ 
c_0^{(R3)} \approx  c  \, \frac{\tilde g_5^2}{g_5^2+ \tilde
g_5^2}~, \ \ 
c_0^{(B)} \approx c  \, \frac{g_5 \, \tilde g_5}{g_5^2+ \tilde
g_5^2} ~.
\eeq
Using these results it can
be checked that the usual SM
relations are exactly satisfied (to leading-log order) and, from the coupling of
the photon and the $W$, we can identify the 4D SM
couplings in terms of the 5D gauge couplings by
\beq
g^2&=&
  \frac{g_5^2\psi^{(L\pm)}(R)^2}{ \int_R^{R^\prime} dz \, \left(
\frac{R}{z}\right) (\psi^{(L\pm)}(z)^2+ \psi^{(R\pm)}(z)^2)}=
\frac{g_5^2}{R \log(R^\prime/R)},
\\
e^2&=&
\frac{\tilde g_5^2 g_5^2 a_0^2}{ \int_R^{R^\prime} dz \, \left(
\frac{R}{z}\right) (2 \tilde g_5^2 +g_5^2)a_0^2}= \frac{g_5^2 \tilde
g_5^2}{(g_5^2+ 2\tilde g_5^2)R \log(R^\prime/R)}.
\eeq
The full SM structure of the $Z$ couplings are also reproduced since:
\beq
\hspace*{-1cm}g^2 \cos \theta_W^2
&=&
\frac{g_5^2\psi^{(L3)}(R)^2}{ \int_R^{R^\prime} dz \, 
\left(\frac{R}{z}\right) 
({\psi^{(L3)}}^2+{\psi^{(R3)}}^2+{\psi^{(B)}}^2)}
=
\frac{g_5^2}{R \log(R^\prime/R)}\, \frac{g_5^2+ \, \tilde
g_5^2}{g_5^2+ 2\tilde g_5^2},
\label{SM1}
\\
g^{\prime 2}&=&
\frac{ g_5^2 \tilde g_5^2}{(g_5^2+\tilde g_5^2) R \log(R^\prime/R)},
\label{SM2}
\\
\sin \theta_W &=& \frac{ \tilde g_5}{\sqrt{ g_5^2+2 \tilde g_5^2} }=
\frac{ g^\prime}{\sqrt{ g^2+  g^{\prime 2}} }.
\label{SM3}
\eeq
Hence the $\rho$ parameter in the leading log approximation is
\begin{equation}
\rho=\frac{M_W^2}{M_Z^2 \cos^2 \theta_W}\approx 1 \ .
\end{equation}
Note, that the fact that the full structure of the SM coupling is 
reproduced implies that at the leading log level there is no 
$S$-parameter either. An $S$-parameter in this language would have manifested
itself in an overall shift of the coupling of the Z compared to its
SM value evaluated from the $W$ and $\gamma$ couplings, which according 
to (\ref{SM1})--(\ref{SM3}) are absent at this order of approximation.
The corrections to the SM relations will appear in the next order 
of the log expansion, and are expected to be of the order of a percent. 
To evaluate the predictions of this model to a precision required by the 
measurements of the electroweak observables one needs to calculate at least 
the next order of corrections to the masses and couplings, 
together with the loop effects of the 
KK gauge bosons, and subtract the usual Higgs contributions.

The next issue is: what are  the masses of the KK excitations of the
$W$ and $Z$?
One can see by numerically solving Eqs.~(\ref{Wtower}) and (\ref{Ztower})
(and using $R= 10^{-19}$ GeV$^{-1}$, $R^\prime= 2 \cdot 10^{-3} {\rm GeV}^{-1}$)
that  $M^W_2 \sim M^Z_2 \sim M^\gamma_2 \sim
1.2$ TeV. These masses are high enough to have evaded direct detection at the Tevatron, but should be within the reach of the LHC.  In terms of an energy expansion, 
the $E^4$ terms of the longitudinal $WW$ scattering would blow up
at energies of few hundred GeV in the 
absence of a Higgs doublet, however to cancel those the effective
four-point vertex obtained from integrating out a heavy $W'$ and $Z'$ is
sufficient. The $E^2$ amplitudes would blow up at $1.8$ TeV, which can
be unitarized by the appearance of these new states. The next set of resonances
arise at $M^W_3 \sim M^Z_3 \sim 1.9$ TeV. 

In the SM the Higgs is used not only to break electroweak
symmetry, but also to generate fermion masses. For technicolor theories
this generically poses a serious problem. In this model, the fermions
can be added as bulk fermions that are doublets of $SU(2)_L$ (the left
handed fermions) and of $SU(2)_R$ (the right handed fermions of the SM).
Bulk fermions are generically Dirac fermions, however on an interval
in warped space only one of the chiralities will have a zero
mode~\cite{inprogress}. The location of the zero mode in warped space
depends on the bulk mass term~\cite{bulkfermions}, and can be localized
close to the Planck brane for the first two generations and the third
generation leptons, which will imply that the gauge couplings for these
fields will be as assumed above. For the right-handed top quark, one can
localize the wave function of the zero mode closer to the TeV brane.

Since the theory on the
TeV brane is vector-like (only $SU(2)_D$ is unbroken there),
a mass for the zero modes can be added
on the TeV brane, which corresponds to a dynamical isospin symmetric
fermion mass in the CFT language. The size of the physical mass will then
depend on the location of the zero mode and the value of the mass
term on the TeV brane.
However because of
the unbroken $SU(2)_D$ symmetry on the TeV brane these masses must be
isospin symmetric, that is the mass for the up and
down type quarks (and similarly for the charged leptons and neutrinos)
are equal at this point. Isospin splitting can be introduced for the leptons
via Majorana masses on the Planck brane  for the right handed neutrinos
using the see-saw mechanism, and
via Dirac mass mixing with extra $SU(2)_R$ singlet fermions on the Planck
brane (isospin breaking can be introduced there since $SU(2)_R$ is broken on
that brane).
For the quarks this will effectively yield a top-quark see-saw type model
for the mass spectrum.

In summary, we have presented a 5D model in warped space where
electroweak symmetry is broken by boundary conditions. The leading
order predictions for the mass spectrum and coupling of the gauge bosons
agree with the SM results, and the first excited $W$ and $Z$ fields
appear at around a  TeV, which is low enough to unitarize the
scattering amplitudes. This model can be viewed as the AdS dual of
a walking technicolor-like theory, and as such one needs to calculate the
leading corrections to electroweak precision observables, which are
estimated to be of order of a percent.

\section*{Acknowledgments}
We thank Kaustubh Agashe, Sekhar Chivukula, Andy Cohen,
Jay Hubisz, Ami Katz, Markus Luty, Riccardo Rattazzi,
Yuri Shirman, Liz Simmons and Raman Sundrum
for useful discussions and comments. We thank the Aspen Center for
Physics  for its hospitality to all four of us while this work was in
progress. C.C. also thanks the T-8 group at Los Alamos for their
hospitality while working on this project.
The research of C.C.
is supported in part by the DOE OJI grant DE-FG02-01ER41206 and in part
by
the NSF grant PHY-0139738.
C.G. and L.P. are supported in part by the RTN European Program
HPRN-CT-2000-00148
and the ACI Jeunes Chercheurs 2068.
J.T. is supported by the US Department of Energy under contract
W-7405-ENG-36. C.C. and C.G. also thank the KITP at UC Santa Barbara for its hospitality while finishing this project. The research done at the KITP is supported by the NSF under 
grant PHY99-07949.



\begin{thebibliography}{99}


\bibitem{CGMPT}
C.~Cs\'aki, C.~Grojean, H.~Murayama, L.~Pilo and J.~Terning,
{\tt hep-ph/0305237}.

\bibitem{otherunitarity}
R.~S.~Chivukula, D.~A.~Dicus and H.~J.~He,
Phys.\ Lett.\ B {\bf 525}, 175 (2002)
[{\tt hep-ph/0111016}];
R.~S.~Chivukula, D.~A.~Dicus, H.~J.~He and S.~Nandi,
{\tt hep-ph/0302263};
R.~S.~Chivukula and H.~J.~He,
Phys.\ Lett.\ B {\bf 532}, 121 (2002)
[{\tt hep-ph/0201164}];
S.~De Curtis, D.~Dominici and J.~R.~Pelaez,
Phys.\ Lett.\ B {\bf 554}, 164 (2003)
[{\tt hep-ph/0211353}];
and Phys.\ Rev.\ D {\bf 67}, 076010 (2003)
[{\tt hep-ph/0301059}];
Y.~Abe, N.~Haba, Y.~Higashide, K.~Kobayashi and M.~Matsunaga,
{\tt hep-th/0302115}.

\bibitem{SonStephanov}
D.~T.~Son and M.~A.~Stephanov,
{\tt hep-ph/0304182}.

\bibitem{Schwartz}
N.~Arkani-Hamed, H.~Georgi and M.~D.~Schwartz,
Annals Phys.\  {\bf 305}, 96 (2003)
{\tt [hep-th/0210184]};
N.~Arkani-Hamed and M.~D.~Schwartz,
{\tt hep-th/0302110};
M.~D.~Schwartz,
{\tt hep-th/0303114}.


\bibitem{RS}
L.~Randall and R.~Sundrum, 
Phys.\ Rev.\ Lett.\  {\bf 83}, 4690 (1999)
{\tt [hep-th/9906064]};
Phys.\ Rev.\ Lett.\  {\bf 83}, 3370 (1999)
{\tt [hep-ph/9905221]}.

\bibitem{holography}
N.~Arkani-Hamed, M.~Porrati and L.~Randall,
JHEP {\bf 0108}, 017 (2001)
{\tt [hep-th/0012148]};
R.~Rattazzi and A.~Zaffaroni,
JHEP {\bf 0104}, 021 (2001)
{\tt [hep-th/0012248]};
M.~Perez-Victoria,
JHEP {\bf 0105}, 064 (2001)
{\tt [hep-th/0105048]};
K.~Agashe and A.~Delgado,
Phys.\ Rev.\ D {\bf 67}, 046003 (2003)
{\tt [hep-th/0209212]}.

\bibitem{KaustubhRaman}
K.~Agashe, A.~Delgado, M.~May 
and R.~Sundrum, {\tt hep-ph/0308036}; K.~Agashe, Cornell Theory Seminar,
May 14, 2003; K.~Agashe, Seminar at the Aspen Center for Physics, July, 2003.


\bibitem{ADS}
K.~Agashe, A.~Delgado and R.~Sundrum,
Annals Phys.\  {\bf 304}, 145 (2003)
{\tt [hep-ph/0212028]}.


\bibitem{CET}
C.~Cs\'aki, J.~Erlich and J.~Terning, Phys.\ Rev.\ D {\bf 66}, 064021
(2002)
{\tt [hep-ph/0203034]}.



\bibitem{RSbulk}
H.~Davoudiasl, J.~L.~Hewett and T.~G.~Rizzo,
Phys.\ Lett.\ B {\bf 473}, 43 (2000)
{\tt [hep-ph/9911262]};
Phys.\ Rev.\ D {\bf 63}, 075004 (2001)
{\tt [hep-ph/0006041]};
A.~Pomarol,
Phys.\ Lett.\ B {\bf 486}, 153 (2000)
{\tt [hep-ph/9911294]};
Phys.\ Rev.\ Lett.\  {\bf 85}, 4004 (2000)
{\tt [hep-ph/0005293]};
S.~Chang, J.~Hisano, H.~Nakano, N.~Okada and M.~Yamaguchi,
Phys.\ Rev.\ D {\bf 62}, 084025 (2000)
{\tt [hep-ph/9912498]};
S.~J.~Huber and Q.~Shafi,
Phys.\ Rev.\ D {\bf 63}, 045010 (2001)
{\tt [hep-ph/0005286]};
L.~Randall and M.~D.~Schwartz,
Phys.\ Rev.\ Lett.\  {\bf 88}, 081801 (2002)
{\tt [hep-th/0108115]};
JHEP {\bf 0111}, 003 (2001)
{\tt [hep-th/0108114]};
S.~J.~Huber, C.~A.~Lee and Q.~Shafi,
Phys.\ Lett.\ B {\bf 531}, 112 (2002)
{\tt [hep-ph/0111465]};
J.~L.~Hewett, F.~J.~Petriello and T.~G.~Rizzo,
JHEP {\bf 0209}, 030 (2002)
{\tt [hep-ph/0203091]};
H.~Davoudiasl, J.~L.~Hewett and T.~G.~Rizzo,
{\tt hep-ph/0212279};
M.~Carena, E.~Ponton, T.~M.~Tait and C.~E.~Wagner,
Phys.\ Rev.\ D {\bf 67}, 096006 (2003)
{\tt [hep-ph/0212307]};
M.~Carena, A.~Delgado, E.~Ponton, T.~M.~Tait and C.~E.~Wagner,
{\tt hep-ph/0305188}.

\bibitem{walking}B. Holdom,  Phys. Rev. {\bf D24} 1441  (1981);
B. Holdom, Phys. Lett. {\bf B150} 301 (1985);
K. Yamawaki, M. Bando, and K. Matumoto, Phys. Rev. Lett.
{\bf 56} 1335  (1986);
T. Appelquist, D. Karabali, and L.C.R. Wijewardhana, Phys. Rev.
Lett. {\em 57} 957  (1986);
T. Appelquist and L.C.R. Wijewardhana, Phys. Rev. {\bf D35} 774  (1987);
T. Appelquist and L.C.R. Wijewardhana, Phys. Rev. {\bf D36} 568 (1987).



\bibitem{A5Higgs}
C.~Cs\'aki, C.~Grojean and H.~Murayama,
Phys.\ Rev.\ D {\bf 67}, 085012 (2003)
[{\tt hep-ph/0210133}];
%
I.~Gogoladze, Y.~Mimura and S.~Nandi,
Phys.\ Lett.\ B {\bf 560}, 204 (2003)
{\tt [hep-ph/0301014]};
Phys.\ Lett.\ B {\bf 562}, 307 (2003)
{\tt [hep-ph/0302176]};
C.~A.~Scrucca, M.~Serone and L.~Silvestrini,
{\tt hep-ph/0304220}.

\bibitem{ST}
M.~E.~Shaposhnikov and P.~Tinyakov,
Phys.\ Lett.\ B {\bf 515}, 442 (2001)
{\tt [hep-th/0102161]}.


\bibitem{inprogress}
C.~Cs\'aki, C.~Grojean, J.~Hubisz, Y.~Shirman and J.~Terning,
to appear.

\bibitem{bulkfermions}
Y.~Grossman and M.~Neubert,
Phys.\ Lett.\ B {\bf 474}, 361 (2000)
{\tt [hep-ph/9912408]};
T.~Gherghetta and A.~Pomarol,
Nucl.\ Phys.\ B {\bf 586}, 141 (2000)
{\tt [hep-ph/0003129]};
S.~J.~Huber and Q.~Shafi,
Phys.\ Lett.\ B {\bf 498}, 256 (2001)
{\tt [hep-ph/0010195]};
Phys.\ Lett.\ B {\bf 512}, 365 (2001)
{\tt [hep-ph/0104293]};
Phys.\ Lett.\ B {\bf 544}, 295 (2002)
{\tt [hep-ph/0205327]}.










\end{thebibliography}
\end{document}